\documentclass[aps,prd,onecolumn,superscriptaddress,preprintnumbers,nofootinbib]{revtex4}%
\usepackage{amsfonts}
\usepackage{amsmath}
\usepackage{graphicx}
\usepackage{xcolor}
\usepackage{wrapfig}
\usepackage[hypertexnames=false]{hyperref}
\usepackage{enumerate}
\usepackage{amssymb}%
\providecommand{\U}[1]{\protect\rule{.1in}{.1in}}
\tighten
\def\cJ{\mathcal{J}}

\def\cP{\mathcal{P}}

\newcommand{\be}{\begin{eqnarray}}
\newcommand{\en}{\end{eqnarray}}
\newcommand{\badat}{\begin{alignedat}}
\newcommand{\eadat}{\end{alignedat}}
\newcommand{\bitm}{\begin{itemize}}
\newcommand{\eitm}{\end{itemize}}
\newcommand{\bmat}{\begin{pmatrix}}
\newcommand{\emat}{\end{pmatrix}}
\newcommand{\ba}{\begin{align}}
\newcommand{\ab}{\end{align}}
\newcommand{\bse}{\begin{subequations}}
\newcommand{\ese}{\end{subequations}}

\def\ndelta{\delta\hspace{-0.50em}\slash\hspace{-0.05em} }

\begin{document}

\title{Asymptotic symmetries and dynamics of three-dimensional flat supergravity} 
\author{Glenn Barnich} \email{gbarnich-at-ulb.ac.be}
\affiliation{Physique Th\'eorique et Math\'ematique Universit\'e Libre
de Bruxelles and International Solvay Institutes Campus Plaine
C.P. 231, B-1050 Bruxelles, Belgium.} 
\author{Laura Donnay}
\email{ldonnay-at-ulb.ac.be} \affiliation{Physique Th\'eorique et
Math\'ematique Universit\'e Libre de Bruxelles and International
Solvay Institutes Campus Plaine C.P. 231, B-1050 Bruxelles, Belgium.}
\author{Javier Matulich} \email{matulich-at-cecs.cl}
\affiliation{Centro de Estudios Cient\'ificos (CECs), Casilla 1469,
Valdivia, Chile.}  \affiliation{Departamento de F\'isica, Universidad
de Concepci\'on, Casilla, 160-C, Concepci\'on, Chile.}
\author{Ricardo Troncoso} \email{troncoso-at-cecs.cl}
\affiliation{Centro de Estudios Cient\'ificos (CECs), Casilla 1469,
Valdivia, Chile.}  \affiliation{Universidad Andr\'es Bello,
Av. Rep\'ublica 440, Santiago, Chile.}  \preprint{CECS-PHY-14/02}

\begin{abstract}
  A consistent set of asymptotic conditions for the simplest
  supergravity theory without cosmological constant in three
  dimensions is proposed. The canonical generators associated to the
  asymptotic symmetries are shown to span a supersymmetric extension
  of the BMS$_{3}$ algebra with an appropriate central charge. The
  energy is manifestly bounded from below with the ground state
  given by the null orbifold or Minkowski spacetime for periodic,
  respectively antiperiodic boundary conditions on the
  gravitino. These results are related to the corresponding ones in
  AdS$_{3}$ supergravity by a suitable flat limit. The analysis is
  generalized to the case of minimal flat supergravity with additional
  parity odd terms for which the Poisson algebra of canonical
  generators form a representation of the super-BMS$_{3}$ algebra with
  an additional central charge.
\end{abstract}

\maketitle

\section{Introduction}

When restricting the gravitational phase-space to conical spacetimes
\cite{Deser:1983tn,Deser:1983dr} in $2+1$-dimensional flat
supergravity, it has been shown \cite{Henneaux:1984ei,Deser:1985qs}
that one can define neither linear momentum nor supercharge but only
energy and angular momentum because the asymptotic dynamics does not
allow for the associated symmetries. The absence of unbroken
supercharge in this context has important physical implications as it
can serve as a mechanism to ensure vanishing cosmological constant for
the vacuum while at the same time boson and fermion masses need no
longer be degenerate \cite{Witten:1994cga}.

The same kind of symmetry breaking occurs in pure Einstein gravity
with negative cosmological constant for a suitably restricted
phase-space, but disappears when consistently relaxing the boundary
conditions in order to allow for a richer asymptotic structure
\cite{Brown:1986nw}: in this case, the asymptotic symmetry group is
enlarged and contains not only $\mathrm{SO}(2,2)$ but the conformal
group in two dimensions. At the same time, the phase-space now
includes, besides the angular defects, further \textquotedblleft zero
mode solutions\textquotedblright, such as the BTZ black hole
\cite{Banados:1992wn,Banados:1993gq} and more generally, two arbitrary
functions that make up the coadjoint representation
\cite{Navarro-Salas1999,Nakatsu1999} of two copies of the Virasoro
algebra at central charge $c^{\pm}=3l/2G$ (see also
\cite{Maloney:2007ud,Garbarz:2014kaa,Barnich:2014zoa} for more recent
discussions and applications). The results on an enhanced asymptotic
structure have been extended to AdS$_3$ supergravity for which the
boundary dynamics is governed by the superconformal algebra
\cite{Coussaert:1994jp,Banados:1998pi,Henneaux:1999ib}.

A similarly rich asymptotic structure for flat three-dimensional
gravity can be defined at null infinity
\cite{Ashtekar1997,Barnich:2006av,Barnich:2010eb} and is connected
through a well-defined flat-space limit to the one of AdS$_3$
\cite{Bagchi:2012cy,Barnich:2012aw,Gonzalez:2013oaa,%
  Costa:2013vza,Fareghbal:2013ifa,Krishnan:2013wta}. In particular,
the limit of BTZ black holes are cosmologies
\cite{Cornalba:2002fi,Cornalba:2003kd} whose horizon entropy can be
understood from symmetry arguments \cite{Barnich:2012xq,Bagchi:2012xr}
consistent with those of the AdS$_3$ case \cite{Strominger:1997eq},
while the boundary dynamics \cite{Coussaert:1995zp} is a flat limit of
Liouville theory \cite{Barnich:2012rz}, obtained through a Hamiltonian
reduction from a chiral $\mathrm{ISO}(2,1)$ Wess-Zumino-Witten theory
\cite{Salomonson:1989fw,Barnich:2013yka}.

The purpose of the present paper is to extend this asymptotic analysis
to the simplest $\mathcal{N}=1$ flat supergravity in three
dimensions. As expected from the AdS and the flat case in the absence
of fermions \cite{Barnich2014}, the reduced phase space with its Dirac
bracket of charges turns out to coincide with the coadjoint
representation of the centrally extended asymptotic symmetry algebra,
viz.~the super-BMS$_{3}$ algebra, which now includes the full
Poincar\'e superalgebra as a subalgebra. 

Note that in the context of Galilean conformal algebras, superalgebras
isomorphic to the super-BMS$_{3}$ algebra, but with a different
physical interpretation for the generators, have been constructed
previously \cite{Bagchi:2009ke,Mandal:2010gx} by taking a
non-relativistic limit of the superconformal algebra (see also
\cite{Sakaguchi:2009de,deAzcarraga:2009ch} for finite-dimensional
versions).

In the next section, we briefly describe $\mathcal{N}=1$ flat
supergravity in three dimensions together with its Chern-Simons
formulation. Additional conventions are given in the appendix
\ref{conventions}.

The main part of the paper is section \ref{Super-BMS}, where we
provide suitable fall-off conditions and work out the asymptotic
symmetry algebra, the general solution to the supergravity equations
of motion consistent with the boundary conditions, the transformation
laws of the functions parametrizing solution space and the Poisson
bracket algebra of the canonical symmetry generators together with the
associated central charge.

Finally, in section \ref{Energy bounds}, we discuss energy bounds and the
Killing spinor equation, while section \ref{Limit of ADS3} is devoted
to rederiving the flat space results from the corresponding ones for
asymptotically AdS$_3$ supergravity by rephrasing the latter in a
suitable gauge that allows one to perform the vanishing cosmological
constant limit in a straightforward way. Section \ref{Reloaded flat
  supergravity} is devoted to the minimal locally supersymmetric
extension of the most general three-dimensional gravity theory without
cosmological constant that leads to first order field equations for
the dreibein and the spin connection. Due to additional parity odd
terms, the Poisson algebra of canonical generators is given again by
the centrally extended super-BMS$_3$ algebra, but now with an
additional central charge for the superrotation subalgebra.

\section{Minimal $\mathcal{N}=1$ flat supergravity in 3D}

\label{Standard N=1 supergravity}

The minimal locally supersymmetric extension of General Relativity in
three dimensions with $\mathcal{N}=1$ gravitino was constructed in
\cite{Deser:1982sw,Deser:1984aa,Marcus:1983hb}. Nowadays, it is
well-known that the theory can be described in terms of a Chern-Simons
action in the cases of negative \cite{Achucarro:1987vz} or vanishing
\cite{Achucarro:1989gm} cosmological constant. In the latter case,
different extensions of the theory have been developed in e.g.,
\cite{Nishino:1991sr,Howe:1995zm,Banados:1996hi,%
  Giacomini:2006dr,Gupta:2007th,Fierro:2014lka}.

Let us begin by considering the simplest case which corresponds to
$\mathcal{N}=1$ supergravity theory with vanishing cosmological
constant. The gauge field $A=A_{\mu}dx^{\mu}$ is given by%
\begin{equation}
A=e^{a}P_{a}+\omega^{a}J_{a}+\psi^{\alpha}Q_{\alpha}\ ,
\end{equation}
where $e^{a}$, $\omega^{a}$ and $\psi^{\alpha}$ stand for the dreibein, the
dualized spin connection $\omega_{a}=\frac{1}{2}\epsilon_{abc}\omega^{bc}$,
and the (Majorana) gravitino, respectively; while the set $\{P_{a}%
,J_{a},Q_{\alpha}\}$ spans the super-Poincar\'{e} algebra, given by%
\[
\lbrack J_{a},J_{b}]=\epsilon_{abc}J^{c}\quad;\quad\lbrack J_{a}%
,P_{b}]=\epsilon_{abc}P^{c}\quad;\quad\lbrack P_{a},P_{b}]=0\ ,
\]%
\begin{equation}
\lbrack J_{a},Q_{\alpha}]=\frac{1}{2}\left(  \Gamma_{a}\right)  _{\hspace
{2mm}\alpha}^{\beta}Q_{\beta}\quad;\quad\lbrack P_{a},Q_{\alpha}]=0\quad
;\quad\{Q_{\alpha},Q_{\beta}\}=-\frac{1}{2}\left(  C\Gamma^{a}\right)
_{\alpha\beta}P_{a}, \label{Super Poincare}%
\end{equation}
where $C$ is the charge conjugation matrix. The action then reads%
\begin{equation}
I[A]=\frac{k}{4\pi}\int\langle A,dA+\frac{2}{3}A^{2}\rangle\ , \label{ICS}%
\end{equation}
where the bracket $\langle\cdot,\cdot\rangle$ stands for an invariant
nondegenerate bilinear form, whose only nonvanishing components are given by
\begin{equation}
\langle P_{a},J_{b}\rangle=\eta_{ab},\quad\langle Q_{\alpha},Q_{\beta}%
\rangle=C_{\alpha\beta}\ , \label{Bracket}%
\end{equation}
and the level is related to the Newton constant according to $k=\frac{1}{4G}$.
Hence, up to a boundary term, the action reduces to%
\begin{equation}
I=\frac{k}{4\pi}\int 2R^{a}e_{a}-\bar{\psi}D\psi\ , \label{Standard action}%
\end{equation}
where $\bar{\psi}_{\alpha}=C_{\alpha\beta}\psi^{\beta}$ is the Majorana
conjugate, while the curvature two-form and the covariant derivative of the
gravitino are defined as%
\begin{equation}
R^{a}=d\omega^{a}+\frac{1}{2}\epsilon^{abc}\omega_{b}\omega_{c}\quad;\quad
D\psi=d\psi+\frac{1}{2}\omega^{a}\Gamma_{a}\psi\ .
\end{equation}
By construction, the action is invariant, up to a surface term,
under the local supersymmetry transformations spanned by $\delta
A=d\lambda+\left[ A,\lambda\right] $, with $\lambda
=\epsilon^{\alpha}Q_{\alpha}$, whose components read%

\begin{equation}
\delta e^{a}=\frac{1}{2}\bar{\epsilon}\Gamma^{a}\psi\ \ ;\ \ \delta\omega
^{a}=0\ \ ;\ \ \delta\psi=D\epsilon\ . \label{SUSY-transf}%
\end{equation}
Analogously, the field equations $F=dA+A^{2}=0$, whose general
solution is locally given by $A=G^{-1}dG$, reduce to%
\begin{equation}
R^{a}=0\ \ ;\ \ T^{a}=-\frac{1}{4}\bar{\psi}\Gamma^{a}\psi\ \ ;\ \ D\psi=0\ ,
\label{Feqs}%
\end{equation}
where $T^{a}=de^{a}+\epsilon^{abc}\omega_{b}e_{c}$ is the torsion
two-form.

\section{Asymptotic behaviour, canonical generators \newline and super-BMS$_{3}$ algebra}

\label{Super-BMS}

The aim is now to provide a suitable set of fall-off conditions for
the gauge fields at infinity that (i)~extends the one of the purely
gravitational sector so as to include the bosonic solutions of
interest, and (ii)~is relaxed enough so as to enlarge the set of
asymptotic symmetries from BMS$_{3}$ to a minimal supersymmetric
extension thereof. In order to fulfill these requirements, the
behaviour of the gauge fields at the boundary is taken to be of the
form%
\begin{equation}
A=h^{-1}a h+h^{-1}dh\ , \label{A}%
\end{equation}
where the radial dependence is completely captured by the group
element $h=e^{-rP_{0}}$, and\footnote{Hereafter we assume light-cone
  coordinates in tangent space. See appendix \ref{conventions} for the
  $\Gamma$-matrices representation and further conventions.}%
\begin{equation}
a=\left(  \frac{\mathcal{M}}{2}du+\frac{\mathcal{N}}{2}%
d\phi\right) P_{0}+du\,P_{1}+\frac{\mathcal{M}}{2}d\phi J_{0}+d\phi\, J_{1}%
+\frac{\psi}{2^{1/4}}d\phi\, Q_{+}\ , \label{flat-connection}%
\end{equation}
where the functions $\mathcal{M}$, $\mathcal{N}$, and the
Grassmann-valued spinor component $\psi$ are assumed to depend on the
remaining coordinates $u$, $\phi$.

The asymptotic symmetries correspond to the set of gauge
transformations, $\delta A=d\lambda+[A,\lambda]$, that preserves this
behaviour. When applied to the dynamical gauge fields $A_{\phi}$, one
finds that the Lie-algebra-valued parameter $\lambda$ must be of the
form%
\begin{equation}
\lambda=\xi^{a}(u,\phi)P_{a}+\chi^{a}(u,\phi)J_{a}+2^{1/4}\epsilon^{+}%
(u,\phi)Q_{+}+2^{1/4}\epsilon^{-}(u,\phi)Q_{-}\ ,
\end{equation}
with%
\be \badat{5} \xi^{0}(u,\phi) & =\frac{1}{2}\mathcal{N}(u,\phi)
\chi^{1}(u,\phi)+\frac{1}{2}\mathcal{M}(u,\phi)\xi^{1}(u,\phi)
-\xi^{1\prime\prime
}(u,\phi)+\frac{1}{2}\epsilon^{-}(u,\phi)\psi(u,\phi)\\
\xi^{2}(u,\phi)  &  =-\xi^{1\prime}(u,\phi)\\
\chi^{0}(u,\phi) & =\frac{1}{2}\mathcal{M}(u,\phi)\chi^{1}(u,\phi
)-\chi^{1\prime\prime}(u,\phi)\\
\chi^{2}(u,\phi)  &  =-\chi^{1\prime}(u,\phi)\\
\epsilon^{+}(u,\phi) & =\frac{1}{\sqrt{2}}\left( \chi^{1}(u,\phi
  )\psi(u,\phi)-2\epsilon^{-\prime}(u,\phi)\right) \
, \label{epsilon+}%
\eadat \en in terms of functions $\chi^1,\xi^1,\epsilon^-$ of
$u,\phi$ and prime denotes a derivative with respect to
$\phi$.  When applied to the Lagrange multipliers $A_{u}$, $\lambda$
is restricted further to depend only on three arbitrary functions of
the angular coordinate, two bosonic ones $Y(\phi)$, $T(\phi)$, and one
fermionic $\mathcal{E}(\phi)$, 
\begin{equation}
\chi^{1}(u,\phi)=Y(\phi)\quad,\quad\epsilon^{-}(u,\phi)=\mathcal{E}(\phi
)\quad,\quad\xi^{1}(u,\phi)=T(\phi)+uY^{\prime}(\phi)\ , 
\end{equation}
and, at the same time, the field equations are required to hold in the
asymptotic region: the fields $\mathcal{M}$, $\mathcal{N}$, $\psi$
become subject to the conditions
\begin{equation}
\partial_{u}\mathcal{M}=0\quad,\quad\partial_{u}\mathcal{N}=\partial_{\phi
}\mathcal{M}\quad,\quad\partial_{u}\psi=0\ , \label{FeqsMNPsi}%
\end{equation} 
which are trivially solved by
\begin{equation}
\mathcal{M}=\mathcal{M}(\phi)\quad,\quad
\mathcal{N}=\mathcal{J}(\phi)+u\mathcal{M}^{\prime}(\phi)\quad,
\quad\psi=\Psi(\phi)\ . \label{MNPSI}%
\end{equation}
The phase space is thus reduced to three arbitrary functions of the
angular coordinate, $\mathcal{M}$, $\mathcal{J}$, $\Psi$, whose
transformation laws under the asymptotic symmetries are given by%
\be \badat{3} \delta\mathcal{M} &
=Y\mathcal{M}^{\prime}+2Y^{\prime}\mathcal{M}%
-2Y^{\prime\prime\prime}\ ,\\
\delta\mathcal{J} & =Y\mathcal{J}^{\prime}+2Y^{\prime}\mathcal{J}+
T\mathcal{M}^{\prime}+2T^{\prime}\mathcal{M}+
\mathcal{E}\Psi^{\prime}+3\mathcal{E}^{\prime}\Psi
-2T^{\prime\prime\prime}\ ,\\
\delta\Psi & =Y\Psi^{\prime}+\frac{3}{2}
Y^{\prime}\Psi+\frac{1}{2}\mathcal{M}\mathcal{E}-
2\mathcal{E}^{\prime\prime}%
\ . \label{delta}%
\eadat \en

The would-be variation of the canonical generators that corresponds to
the asymptotic symmetries spanned by $\lambda(T,Y,\mathcal{E})$ can be readily
found in the canonical approach \cite{Regge:1974zd}. In the case of a
Chern-Simons theory in three dimensions, they are given by
\cite{Balachandran:1991dw,Banados:1994tn,Carlip:2005zn,Perez:2014pya}
\begin{equation}
  \ndelta Q[\lambda]=-\frac{k}{2\pi}\int\langle\lambda, 
\delta A_{\phi}\rangle d\phi\ .
\end{equation}
For the asymptotic behaviour described here, it is straightforward to
verify that this expression becomes linear in the deviation of the
fields with respect to the reference background, so that it can be
directly integrated as%
\begin{equation}
  Q[T,Y,\mathcal{E}]=-\frac{k}{4\pi}\int\left[  T\mathcal{M}+%
    Y\mathcal{J}-2\mathcal{E}\Psi\right]  d\phi\ .\label{Q-flat}%
\end{equation}
Therefore, since the Poisson brackets fulfill $\delta_{\lambda_{1}}%
Q[\lambda_{2}]=\{Q[\lambda_{2}],Q[\lambda_{1}]\}$, the algebra of the
canonical generators can be directly read from the transformation laws
in (\ref{delta}). When expanded in Fourier modes,
\[
\mathcal{P}_{m}= \frac{k}{4\pi}\int e^{im\phi} \mathcal{M}\, d\phi
\quad , \quad \mathcal{J}_{m} =\frac{k}{4\pi}\int e^{im\phi}
\mathcal{J}\, d\phi \quad , \quad \mathcal{Q}_{m} =
\frac{k}{4\pi}\int e^{im\phi}
\Psi\, d\phi ,%
\]
the Poisson brackets read explicitly

\be \badat{4} 
i \{ \mathcal{P}_{m},\mathcal{P}_{n} \} &
=0 \ , \\
i \{ \mathcal{J}_{m},\mathcal{J}_{n} \} &
=(m-n)\mathcal{J}_{m+n}+\frac{c_{1}} %
{12}m^{3}\delta_{m+n,0} \ , \\
i \{ \mathcal{J}_{m},\mathcal{P}_{n} \} &
=(m-n)\mathcal{P}_{m+n}+\frac{c_{2}}%
{12}m^{3}\delta_{m+n,0} \ , \\
i \{ \mathcal{P}_{m},\mathcal{Q}_{n} \}  &  =
0\ , \\
i \{ \mathcal{J}_{m},\mathcal{Q}_{n} \}  &  =
\left(  \frac{m}{2}-n\right)  \mathcal{Q}_{m+n}\ , \\
\{\mathcal{Q}_{m},\mathcal{Q}_{n}\} &
=\mathcal{P}_{m+n}+\frac{c_{2}}{6}m^{2}\delta_{m+n,0}\ ,
\label{Super-BMS-simplest}
\eadat \en 
where the central charges are given by $c_1=0$ and
$c_{2}=\frac{3}{G}$. Note that the standard redefinitions
$\cJ_0\rightarrow \cJ_0 +\frac{c_1}{24}$, $\cP_0\rightarrow \cP_0
+\frac{c_2}{24}$ change the central terms in the algebra to
$\frac{c_1}{12} m(m^2-1)\delta_{m+n,0}$, $\frac{c_2}{12}
m(m^2-1)\delta_{m+n,0}$ and $\frac{c_2}{6}(m^2-\frac{1}{4})$.

Algebra (\ref{Super-BMS-simplest}) constitutes the minimal
supersymmetric extension of the BMS$_{3}$ algebra with central
extensions. One can furthermore show that the fields
$\mathcal{M},\mathcal{J},\Psi$, their transformation laws
\eqref{delta} and the Poisson bracket algebra
(\ref{Super-BMS-simplest}) are entirely captured by the coadjoint
representation of the centrally extended super-BMS$_3$ group. 

\section{Energy bounds and Killing spinors}

\label{Energy bounds}

\subsection{Energy bounds from quantum superalgebra}
\label{sec:energy-bounds-from}

If the gravitino fulfills antiperiodic (Neveu-Schwarz) boundary
conditions, the modes $\mathcal{Q}_{p}$ involve half-integer $p$. The
wedge subalgebra is then spanned by the subset $\mathcal{P}_{m}$,
$\mathcal{J}_{m}$, $\mathcal{Q}_{p}$, with $m=\pm 1,0$, and
$p=\pm{1}/{2}$, which corresponds to the super-Poincar\'{e} algebra.
Indeed, this can be explicitly seen once the modes in
(\ref{Super-BMS-simplest}) are identified with the generators in
(\ref{Super Poincare}) according to $\mathcal{J}_{-1}=-\sqrt{2}J_{0}$,
$\mathcal{J}_{1}=\sqrt{2}J_{1}$, $\mathcal{J}_{0}=J_{2}$,
$\mathcal{P}%
_{-1}=-\sqrt{2}P_{0}$, $\mathcal{P}_{1}=\sqrt{2}P_{1}$,
$\mathcal{P}_{0}%
=P_{2}-\frac{1}{8G}$, $\mathcal{Q}_{1/2}=\sqrt{2}Q_{-}$ and
$\mathcal{Q}%
_{-1/2}=\sqrt{2}Q_{+}$.  In the quantum theory, one can then use
arguments similar to those of
\cite{Deser:1977hu,Abbott:1981ff,Coussaert:1994jp}: the last of the
brackets in (\ref{Super-BMS-simplest}) becomes an anticommutator to
lowest order in $\hbar$ and the quantum generator $\mathcal{P}_{0}$ is
bounded according to%
\begin{equation}
\mathcal{P}_{0}=\mathcal{Q}_{1/2}\mathcal{Q}_{-1/2}+
\mathcal{Q}_{-1/2}\mathcal{Q}_{1/2}-\frac{1}{8G}\geq-\frac{1}%
{8G}\ . \label{Bound-NS}%
\end{equation}
In classical supergravity, the simplest solution that saturates the
bound is Minkowski spacetime with $\mathcal{P}_{0}=-\frac{1}{8G}$ and
all other modes of $\mathcal{M},\mathcal{J},\Psi$ vanishing.

For the case of periodic (Ramond) boundary conditions for the
gravitino, the modes $\mathcal{Q}_{p}$ involve integer $p$
and the bound on the quantum generator becomes
\begin{equation}
\mathcal{P}_{0}=\mathcal{Q}_{0}^{2}\geq0\ . \label{Bound-Ramond}%
\end{equation}
The simplest classical supergravity solution that saturates this bound
is the null orbifold \cite{Horowitz:1990ap} with all modes
vanishing\footnote{How to turn these arguments into a supersymmetry
  based proof, analogous to the one in four dimensions
  \cite{Witten:1981mf}, of the positive energy theorems in classical
  three-dimensional general relativity \cite{Barnich:2014zoa} will be
  discussed elsewhere.}. 

\subsection{Asymptotic Killing spinors}
\label{sec:asympt-kill-spin}

Starting from transformations \eqref{delta}, one can systematically
discuss the isotropy subalgebras of various solutions. A particular
case of this problem is the \textquotedblleft asymptotic Killing
spinor equation\textquotedblright, i.e., the question which asymptotic
supersymmetry transformations leave purely bosonic solutions
invariant,
\begin{equation}
  \delta_{\mathcal{E}}\Psi=-2\mathcal{E}^{\prime\prime}+\frac{1}{2}\mathcal{M}
\mathcal{E}=0\ . 
\label{Asympt-KS}%
\end{equation}
Asymptotic Killing spinors of solutions with constant
$\mathcal{M}\neq0$, are given by 
\begin{equation}
\mathcal{E}=Ae^{\frac{\sqrt{\mathcal{M}}}{2}\phi}+Be^{-\frac{\sqrt{\mathcal{M}}%
}{2}\phi}\ , \label{epsilon generic}%
\end{equation}
with $A$, $B$ constants. They are globally well-defined provided
$\mathcal{M}=-n^{2}$, with $n> 0$ a strictly positive integer, 
\begin{equation}
  \mathcal{E}=\mathcal{E}_{\frac{n}{2}}e^{in\frac{\phi}{2}}+
  \mathcal{E}_{-\frac{n}{2}}e^{-in\frac{\phi}{2}}\
  . \label{epsilon windings}
\end{equation}
Solutions with
$n>1$ are below the bounds (\ref{Bound-NS}) or
(\ref{Bound-Ramond}). This singles out $n=1$, Minkowski spacetime for
$\mathcal{J}=0$, in which case there are
two independent antiperiodic solutions. 

In the remaining case, $\mathcal{M}=0$, the solution of (\ref{Asympt-KS}) is
given by%
\begin{equation}
\mathcal{E}=\mathcal{E}_{0}+\mathcal{F}_{0}\, \phi\ ,\label{epsilon M=0}%
\end{equation}
with $\mathcal{E}_{0}$, $\mathcal{F}_{0}$ constants, which is
single-valued provided $\mathcal{F}_{0}=0$. This means in particular that there
is a single periodic solution for the null orbifold at
$\mathcal{J}=0$.

\subsection{Exact Killing spinors of bosonic zero mode solutions}
\label{sec:exact-kill-spin}

Purely bosonic solutions ($\psi=0$) to the field equations
(\ref{Feqs}) in the asymptotic region are described in outgoing
Eddington-Finkelstein coordinates by metrics%
\begin{equation}
ds^{2}=\mathcal{M}du^{2}-2dudr+\mathcal{N}dud\phi+r^{2}d\phi^{2}\ ,
\label{Metric}%
\end{equation}
with $\mathcal{M},\mathcal{N}$ as in \eqref{MNPSI}. 
The ``zero mode solutions''
\begin{equation}
\mathcal{M}=8GM\quad,\quad\mathcal{N}=8GJ\ ,\label{constants}
\end{equation}
with $M,J$ constants, describe cosmological solutions for nonnegative
mass ($M\geq0$) and arbitrary values of the angular momentum $J$,
while for $-\frac{1}{8G}<M<0$, the geometry corresponds to stationary
conical defects. For $M=-\frac{1}{8G}$, the curvature is no longer
singular at the origin, but the torsion is unless $J=0$, which
corresponds to Minkowski spacetime. Below this value of the mass, the
geometry describes angular excesses (see, e.g.,
\cite{Deser:1983tn,Barnich:2012aw}).

\bigskip

Such solutions admit global supersymmetries when they
are invariant under supersymmetry transformations of the form
(\ref{SUSY-transf}), provided the spinorial parameter $\epsilon$ is globally
defined. The Killing spinor equation to be solved is then given by%
\begin{equation}
D\varepsilon=(d+\omega)\varepsilon=0\ ,\label{KS-eq}%
\end{equation}
with $\omega=\frac{1}{2}\omega^{a}\Gamma_{a}$. 

This equation can be solved directly through
$\varepsilon=\Lambda^{-1}\varepsilon_{0}$ with $\varepsilon_{0}$ a
constant spinor and $\Lambda$ the Lorentz group element associated to
the flat spin connection, $\omega=\Lambda^{-1}d\Lambda$, whose form
can be read off \eqref{flat-connection},
\[
\Lambda=\exp\left[  \frac{1}{2}\left(  \Gamma_{1}+\frac{1}{2}\mathcal{M}\Gamma
_{0}\right)  \phi\right]  =\left(
\begin{array}
[c]{cc}%
\cosh\left(  \frac{\sqrt{\mathcal{M}}}{2}\phi\right)   & \sqrt{\frac
{\mathcal{M}}{2}}\sinh\left(  \frac{\sqrt{\mathcal{M}}}{2}\phi\right)  \\
\sqrt{\frac{2}{\mathcal{M}}}\sinh\left(  \frac{\sqrt{\mathcal{M}}}{2}%
\phi\right)   & \cosh\left(  \frac{\sqrt{\mathcal{M}}}{2}\phi\right)
\end{array}
\right)  \ .
\]
Alternatively, one can first solve the Killing spinor equation for the
upper component. According to (\ref{epsilon+}), this amounts to
$\epsilon^{+}=-\sqrt{2}\epsilon^{-\prime}$. The equation for the lower
component then reduces to the asymptotic Killing spinor equation
\eqref{Asympt-KS}. 

When suitably identifying the constants $\epsilon_0^+,\epsilon_0^-$,
one finds in both cases that the Killing spinor $\varepsilon$ is
globally defined provided $\mathcal{M}=-n^{2}$ with $n$ a positive
integer. For $n>0$, one finds two independent Killing spinors which
can be periodic (even $n$) or antiperiodic (odd $n$) given explicitly
by $\epsilon=(-\sqrt{2} \mathcal{E}^{\prime},\mathcal{E})$, with $\mathcal{E}$ as in \eqref{epsilon
  windings}. For $n=0$, one finds a single independent periodic
solution given explicitly by $\epsilon=(0,\mathcal{E}_0)$.

In summary, massive cosmological solutions ($\mathcal{M}>0$) do not
admit global supersymmetries, while the massless case admits only one
periodic Killing spinor. For $\mathcal{M}=-n^{2}$, the geometries
possess two (the maximum number of) global supersymmetries, which
includes, for $n=1$, the case of Minkowski spacetime.  \bigskip

Note that the geometries with $\mathcal{M}=-n^{2}$, $n > 1$ can be
interpreted as suitable unwrappings of those for $n=1$ with $n$ playing
the role of the winding number. Indeed, the rescalings
\[
\phi^{\prime}=n\phi\hspace{1mm},\hspace{8mm}r^{\prime}=n^{-1}r\hspace
{1mm},\hspace{8mm}u^{\prime}=nu,
\]
amount to the change $M\rightarrow n^2M$, $J\rightarrow n^2 J$ in
(\ref{constants}). As we
have argued in section \ref{sec:asympt-kill-spin}, these geometries
actually become excluded when one insists on fulfilling the energy
bounds in eqs. (\ref{Bound-NS}) and (\ref{Bound-Ramond}), for the
periodic and antiperiodic boundary conditions, respectively.

It is worth pointing out that geometries endowed with angular deficit
or excess actually possess a curvature singularity on top of the
source at the origin, so that they do not fulfill the integrability
condition of (\ref{KS-eq}), i.e., $DD\varepsilon\neq0$. Minkowski
spacetime is obviously devoid of this problem, while a detailed
discussion of the singularity of the null orbifold
$\mathcal{M}=0=\mathcal{J}$ at $r=0$ can be found in Section 2.3 of
\cite{Liu:2002ft}.

\section{Flat limit of asymptotically AdS$_{3}$ supergravity}

\label{Limit of ADS3}

The standard $\mathcal{N}=1$ supergravity action (\ref{Standard
  action}) can be directly recovered either from the ($1,0$) or the
($0,1$) AdS supergravity theory in the vanishing cosmological constant
limit. However, when one deals with the asymptotic behaviour of the
fields, even in the case of pure gravity, the limiting process turns
out to be much more subtle \cite{Barnich:2012aw}.  In this section we
show how the results obtained in section \ref{Super-BMS} can be
recovered from the corresponding ones in the case of asymptotically
AdS$_{3}$ supergravity. Here we follow a similar strategy as the one
carried out in \cite{Gonzalez:2013oaa} for the vanishing cosmological
constant limit of higher spin gravity, which consists in finding a
particularly suitable gauge choice that allows to perform the limit in
a straightforward way.

\subsection{Asymptotic behaviour of minimal AdS$_{3}$ supergravity,
  canonical generators \newline and superconformal symmetry}

\label{AdS-behaviour}

There are two inequivalent minimal locally supersymmetric extensions
of General Relativity with negative cosmological constant in three
spacetime dimensions, known as the ($1$,$0$) and ($0$,$1$)
theories. Since both possess the same vanishing cosmological limit,
without loss of generality we will choose the ($1$,$0$) one, which can
be formulated as a Chern-Simons theory whose gauge group is given by
$OSp(2|1)\otimes Sp(2)$\ \cite{Achucarro:1987vz}%
. The action depends on two independent connections $A^{+}$ and
$A^{-}$, for $OSp(2|1)$ and $Sp(2)$, respectively, and is given by%
\[
I_{\mathrm{SAdS}}=I[A^{+}]-I[A^{-}]\ ,
\]
where $I[A]$ is defined in (\ref{ICS}).

The asymptotic behaviour of the fields has been previously discussed
in \cite{Banados:1998pi, Henneaux:1999ib}. The fall-off of the
fields can be written as%
\begin{equation}
A^{\pm}=b_{\pm}^{-1}a^{\pm}b_{\pm}+b_{\pm}^{-1}db_{\pm}  \ , \label{A-new gauge}%
\end{equation}
with $b_{\pm}=e^{\pm\log(r/l)L_{0}}$, and%

\be
\badat{2}
a^{+}  &  =\left(  L_{1}^{+}-\mathcal{L}_{+}L_{-1}^{+}+\psi
Q_{+}\right)  dx^{+}\ ,\\
a^{-}  &  =  \left( L_{-1}^{-}-\mathcal{L}_{-}L_{1}^{-}\right)  dx^{-}\ ,
\label{a-ads}%
\eadat \en where $x^{\pm}=\frac{t}{l}\pm\phi$. Here the generators
$L_{i}^{\pm}$, with $i=-1,0,1$, span the left and right copies of
$Sp(2)$, and $Q_{\alpha}$, with $\alpha=1,-1$, correspond to the
(left) fermionic generators of $OSp(2|1)$.  On-shell, the functions
$\mathcal{L}_{\pm}$ and the Grassmann-valued $\psi$, are required to be
chiral, i.e.,
\begin{equation}
\partial_{\mp}\mathcal{L}_{\pm}=0,\quad\partial_{-}\psi=0\
, \label{chirality AdS}%
\end{equation}
so that they depend only on $x^{+}$ or $x^{-}$.

The asymptotic symmetries are given by the gauge transformations $\delta
a^{\pm}=d\lambda^{\pm}+[a^{\pm},\lambda^{\pm}]$ that maintain the form of
(\ref{a-ads}), so that $\lambda^{\pm}$ are given by%

\[
\lambda^{+}=\chi^{+} L_{1} - \chi^{+ \prime} L_0 + \frac{1}{2}\left(
  -2 \mathcal{L}_{+}\chi^{+} - \epsilon \Psi + \chi^{+ \prime \prime}
\right) L_{-1} + (\chi^{+} \Psi + 
  \epsilon^{\prime} ) Q_{+} +  \epsilon Q_{-}
\ ,
\]
and%
\[
\lambda^{-}= \chi^{-} L_{-1} + \chi^{- \prime} L_0 + \frac{1}{2}
\left(-2 \mathcal{L}_{-} \chi^{-} + \chi^{- \prime \prime}
\right)L_{1},
\]
which depend on three arbitrary chiral functions, fulfilling%
\begin{equation}
  \partial_{\pm}\chi^{\mp}=0 \quad \text{,} \quad \partial_{-}
  \epsilon=0\ . \label{chirality parameters}%
\end{equation}
The on-shell transformation law of the fields $\mathcal{L}_{\pm}$,
$\psi$ reads%
\begin{align}
  \badat{3}
  \delta \mathcal{L}_{+} &= \chi^{+} \mathcal{L}_{+}^{ \prime}  + 2
  \mathcal{L}_{+} \chi_{+}^{ \prime} -\frac{1}{2} \chi^{+ \prime
    \prime \prime } + \frac{3}{2} \psi \epsilon^{- \prime} +
  \frac{1}{2} \epsilon^{-} \psi^{\prime},   \\
  \delta \psi &= - \mathcal{L}_{+} \epsilon^{-} +  \epsilon^{-
    \prime \prime} + \frac{3}{2} \psi \chi^{+ \prime} + \chi^{+}
  \psi^{\prime},\\ 
  \delta \mathcal{L}_{-} &= \chi^{-} \mathcal{L}_{-}^{ \prime} + 2
  \mathcal{L}_{-} \chi_{-}^{ \prime} -\frac{1}{2} \chi^{- \prime
    \prime \prime } .
\label{deltafields AdS}
\eadat
\end{align}

The canonical generators associated to the asymptotic symmetries
spanned by $\lambda^{+}=\lambda^{+}(\chi^{+} , \epsilon)$ and
$\lambda^{-}%
=\lambda^{-}(\chi^{-})$, are given by%
\be \badat{2} Q^{+}[\chi^{+} , \epsilon] & =-\frac{\kappa}{2\pi}\int\left[
  \chi^{+}\mathcal{L}_{+} - \epsilon \psi \right]
d\phi\ ,\label{Q+}\\
Q^{-}[\chi^{-}] & =-\frac{\kappa}{2\pi}\int\left[ \chi^{-} \mathcal{L}_{-}
\right] d\phi\ ,\label{Q-}%
\eadat \en where $\kappa : = l k$,  which by virtue of (\ref{deltafields AdS}), can be readily
shown to fulfill the (super) Virasoro algebra. Expanding in Fourier
modes $\mathcal{L}^\pm_{m}=\frac{kl}{4\pi}\int \mathcal{L}_{\pm}
e^{\pm i m \phi}\, d\phi$ and $\mathcal{Q}_{m}=\frac{kl}{4 \pi} \int
\psi e^{i m \phi}\, d\phi$ , the nonvanishing Poisson brackets read
\be \badat{3} i \{ \mathcal{L}_{m}^{\pm},\mathcal{L}_{n}^{\pm} \} &
=(m-n)\mathcal{L}_{n+m}^{\pm}+\frac{c}{12}%
m^{3}\delta_{m+n,0}\ , \\
i \{ \mathcal{L}_{m}^{+},\mathcal{Q}_{n}^{+} \} & =\left(
  \frac{m}{2}-n\right) \mathcal{Q}_{m+n}%
^{+}\ , \\
\{\mathcal{Q}_{m}^{+},\mathcal{Q}_{n}^{+}\} &
=2\mathcal{L}_{m+n}^{+}+\frac{c}{3}m^{2}\delta _{m+n,0} \ .
\label{superconformal algebra}
\eadat
\en
where the central charge is given by $c=\frac{3l}{2G}$. \newline

\subsection{Vanishing cosmological constant limit}

\label{Flat limit}

In order to recover the results of section \ref{Super-BMS} from the
ones described in the previous subsection once the vanishing
cosmological constant limit is taken, it turns out to be useful to
express the asymptotic behaviour of the gauge fields of the ($1$,$0$)
AdS supergravity theory in a different gauge. We then choose different
group elements $g_{\pm}$, so that the fall-off of the connections
now read%
\begin{equation}
A^{\pm}=g_{\pm}^{-1}a^{\pm}g_{\pm}+g_{\pm}^{-1}dg_{\pm}\ ,
\end{equation}
where $a^{\pm}$ are given by (\ref{a-ads}), and%
\be \badat{2} g_{+} &
=b_{+}e^{-\log(\frac{r}{l})L_{0}}e^{\frac{r}{2l}%
  L_{-1}}\ ,\\
g_{-} &
=b_{-}e^{-\log(\frac{r}{4 l})L_{0}}e^{\frac{r}%
  {2 l}L_{-1}}e^{\frac{2 l}{r}L_{1}}\ .\label{g+-}%
\eadat \en In this gauge, the asymptotic form of the super-AdS gauge
field is explicitly given by%
\be \badat{2} A^{+} &
=\frac{r}{l}dx^{+}L_{0}^{+}+\frac{1}{2}\left[ \frac{dr}%
  {l}+\left( \frac{r^{2}}{2l^{2}}-2\mathcal{L}_{+}\right)
  dx^{+}\right] L_{-1}%
^{+}+dx^{+}L_{1}^{+}+\psi^{+}Q_{+}%
dx^{+}\ ,\\
A^{-} & =\frac{r}{l}dx^{-}L_{0}^{-}-\frac{1}{2}\left[
  \frac{dr}%
  {l}+\left( \frac{r^{2}}{2l^{2}}-2\mathcal{L}_{-}\right)
  dx^{-}\right] L_{-1}%
^{-}- dx^{-}L_{1}^{-}\ .  \eadat \en It is now
convenient to make the change $t=u$ and to perform the change of
basis%
\be \badat{2} L_{-1}^{(\pm)}=-2 J_{0}^{\pm},\quad
L_{0}^{\pm}=J_{2}^{\pm}, \quad L_{1}^{(\pm)}= J_{1}^{\pm}\ , 
\quad Q_{+} = \frac{1}{2^{1/4}}\tilde{Q}_{+},
\eadat \en 
followed by%
\begin{equation}
J_{a}^{\pm}=\frac{J_{a}\pm lP_{a}}{2},\quad Q_{+}=\sqrt{l}\tilde{Q}_{+}\ ,
\end{equation}
so that the full gauge field reads%
\begin{eqnarray}
A &=& \left(  -dr+\frac{\mathcal{M}}{2}du+\frac{\mathcal{N}}{2}d\phi -
  \frac{r^2}{2 l^2} du\right)  P_{0}%
+duP_{1}+rd\phi P_{2} +\left(\frac{\mathcal{M}}{2}d\phi +
  \frac{\mathcal{N}}{2l^2} du - \frac{r^2}{2l^2} d\phi\right) J_{0}
\nonumber \\ 
&&+d\phi J_{1} + \frac{r}{l^2} du
J_2+\frac{\Psi}{2^{1/4}}\tilde{Q}_{+}d\phi+\frac{1}{l}\frac{\Psi }{
  2^{1/4}}\tilde{Q}_{+}du \ , 
\end{eqnarray} 
where the arbitrary functions $\mathcal{L}_{\pm}$, $\psi$ have been
redefined according to%
\begin{equation}
\mathcal{M}=(\mathcal{L}_{+}+\mathcal{L}_{-}),\quad
\mathcal{N}=l(\mathcal{L}_{+}-\mathcal{L}_{-}),\quad   
\Psi=\sqrt{l}\psi\ .\label{redef-functions}%
\end{equation}
The chirality conditions (\ref{chirality AdS}) now read%
\begin{equation}
\partial_{u} \mathcal{M} = \frac{1}{l^2} \partial_{\phi} \mathcal{N}
\quad , \quad  \partial_{u} \mathcal{N} = \partial_{\phi} \mathcal{M}  
\quad , \quad \partial_{u} \Psi= \frac{1}{l} \partial_{\phi} \Psi 
.\label{Chirality-AdS-redefined}%
\end{equation}

\bigskip

The vanishing cosmological constant limit can now be directly
performed in a transparent way. Indeed, for the full gauge field
$A=A^{+}+A^{-}$, one just takes $l\rightarrow\infty$, so that it
reduces to%
\begin{equation}
A=\left(
  -dr+\frac{\mathcal{M}}{2}du+ \frac{\mathcal{N}}{2}d\phi\right)P_{0}%
+duP_{1}+rd\phi P_{2}+\frac{\mathcal{M}}{2}d\phi J_{0}+d\phi
J_{1}+\frac{\Psi }{2^{1/4}}Q_{+}d\phi\ ,\nonumber 
\end{equation}
which coincides with the asymptotic form of the connection in the
asymptotically flat case, eqs. (\ref{A}), (\ref{flat-connection}).
Analogously, in the limit, the chirality conditions
(\ref{Chirality-AdS-redefined}) take the flat space form
(\ref{FeqsMNPsi}), whose solution is given by (\ref{MNPSI}).

It is simple to verify that the expression for the global charges for
the gauge choice (\ref{g+-}) remains the same as in the gauge
(\ref{A-new gauge}) and is still given by (\ref{Q+}). After making use
of the redefinition for the fields in (\ref{redef-functions}), they
acquire the form%
\begin{align}
Q [f,h,\mathcal{E}] &  =-\frac{k}{4\pi} \int d \phi \left( f
  \mathcal{M} +  h \mathcal{N} - 2 \mathcal{E} \Psi
\right), \label{Qcontutti}%
\end{align}
where the parameters that characterize the asymptotic symmetries have
been conveniently redefined as%
\[
 f=l(\chi^{+}+\chi
^{-}),\quad h=\chi^{+}-\chi^{-} ,\quad \mathcal{E}  = \sqrt{l} \varepsilon.
\]
The chirality conditions (\ref{chirality parameters}) on the gauge
parameters then read%
\begin{equation}
  \partial_{u}f=\partial_{\phi}h,\quad\partial_{u}h=\frac{1}{l^{2}%
  }\partial_{\phi}f, \quad \partial_{u} \mathcal{E}
  =\frac{1}{l}\partial_{\phi} \mathcal{E}   ,\label{chirality-redef}%
\end{equation}
and, in the limit $l\rightarrow\infty$, they imply that%
\[
h=Y(\phi),\quad f=T(\phi)+u Y^{\prime},\quad \mathcal{E} =
\mathcal{E}(\phi) ,
\]
and hence, by virtue of (\ref{MNPSI}), the global charges
(\ref{Qcontutti}) reduce to the ones of the asymptotically flat case 
given in (\ref{Q-flat}).

\bigskip

As explained in section \ref{AdS-behaviour}, the canonical generators of
($1$,$0$) AdS supergravity satisfy the centrally-extended superconformal
algebra in two dimensions given by (\ref{superconformal algebra}). In
order to take the flat limit, it is useful to change the basis according to%
\[
P_{m}\equiv\frac{1}{l}(L_{m}^{+}+L_{-m}^{-})\ ,\hspace{8mm}J_{m}\equiv
L_{m}^{+}-L_{-m}^{-}\ ,
\]
as well as rescaling the supercharges as%
\[
Q_{m}\equiv\frac{1}{\sqrt{l}}Q_{m}^{+}\ .
\]
After this has been done, in the limit $l\rightarrow\infty$, algebra
(\ref{superconformal algebra}) readily reduces to the minimal
supersymmetric extension of the BMS$_{3}$ algebra
(\ref{Super-BMS-simplest}), where the central charges are given by
$l c_1 = c^{+} - c^{-}$ and $l c_2= c^{+} + c^{-}$. In particular, it also
follows that the bounds for the generators that are obtained from the
superconformal algebra, reduce to the ones in eqs. (\ref{Bound-NS})
and (\ref{Bound-Ramond}) in the limit of vanishing cosmological
constant.

\section{Asymptotic structure of $\mathcal{N}=1$ \textquotedblleft
  reloaded\textquotedblright\ flat supergravity}

\label{Reloaded flat supergravity}

The locally supersymmetric extension of the most general
three-dimensional gravity theory that leads to first order field
equations for the dreibein and the spin connection has been
constructed in \cite{Giacomini:2006dr}. It includes additional
parity-odd terms as compared with the standard theory. In the
vanishing cosmological constant limit, the action with $\mathcal{N}=1$
supersymmetry is given by%
\begin{equation}
I_{(\mu,\gamma)}=\frac{k}{4\pi}\int2\left(  1+\mu\gamma\right)  R^{a}%
e_{a}+\gamma^{2} \left(  1+\mu\frac{\gamma}{3}\right)  \epsilon_{abc}%
e^{a}e^{b}e^{c}+\mu L(\omega)+\gamma\left(  2+\mu\gamma\right)  T^{a}%
e_{a}-\bar{\psi}\left(  D+\frac{\gamma}{2}e^{a}\Gamma_{a}\right)  \psi\ ,
\label{Imugamma}%
\end{equation}
where $L(\omega)=\omega^{a}d\omega_{a}+\frac{1}{3}\epsilon_{abc}\omega
^{a}\omega^{b}\omega^{c}$ is the Lorentz-Chern-Simons form. This
action is invariant, up to a surface term, under the following local
supersymmetry transformations%
\begin{equation}
\delta e^{a}=\frac{1}{2}\bar{\epsilon}\Gamma^{a}\psi\ \ ,\ \
\delta\omega 
^{a}=\frac{1}{2}\gamma\bar{\psi}\Gamma^{a}\epsilon\ \ ,
\ \ \delta
\psi=D\epsilon+\frac{1}{2}\gamma e^{a}\Gamma_{a}\epsilon\ .
\label{susy reloaded}%
\end{equation}
Note that in the case of $\mu=\gamma=0$, the action (\ref{Imugamma})
and the supersymmetry transformations (\ref{susy reloaded}) reduce to
the standard ones, given by (\ref{Standard action}) and
(\ref{SUSY-transf}), respectively.  

Remarkably, in spite of the presence of a volume term in
(\ref{Imugamma}), the theory can also be formulated in terms of a
Chern-Simons action for the super-Poincar\'{e} group. This can be seen 
as follows. In terms of the shifted spin connection
$\hat{\omega}^{a}:=\omega^{a}+\gamma e^{a}$, action (\ref{Imugamma})
reads%
\begin{equation}
I_{(\mu,\gamma)}=\frac{k}{4\pi}\int2\hat{R}^{a}e_{a}+\mu L(\hat{\omega}%
)-\psi_{\alpha}\hat{D}\psi^{\alpha}\ ,
\end{equation}
where $\hat{D}$, $\hat{R}^{a}$, and $L(\hat{\omega})$ stand for the
covariant derivative, the curvature two-form, and the
Lorentz-Chern-Simons form constructed out from $\hat{\omega}^{a}$,
respectively. Hence, up to a boundary term, the action can be written
as%
\begin{equation}
I[A]=\frac{k}{4\pi}\int\langle A,dA+\frac{2}{3}A^{2}\rangle\ ,
\end{equation}
where now the gauge field is given by%
\begin{equation}
A=e^{a}P_{a}+\hat{\omega}^{a}J_{a}+\psi^{\alpha}Q_{\alpha}\ ,
\label{A reloaded}%
\end{equation}
and the nonvanishing components of the invariant nondegenerate
bilinear form 
read%
\begin{equation}
\langle P_{a},J_{b}\rangle=\eta_{ab},\quad\langle J_{a},J_{b}\rangle=\mu
\eta_{ab},\quad\langle Q_{\alpha},Q_{\beta}\rangle=C_{\alpha\beta}\ ,
\label{CS3}%
\end{equation}
so that it reduces to the standard bracket in (\ref{Bracket}) in the
case of $\mu=0$.

\bigskip

The asymptotic behaviour of the gauge fields in this case is then
proposed to be exactly of the same form as in eqs. (\ref{A}),
(\ref{flat-connection}), which by virtue of (\ref{A reloaded}),
amounts just to modify the fall-off of the spin connection
$\omega^{a}$ in the asymptotic region. This has to be so because the
field equations now imply a nonvanishing torsion even in vacuum.

Therefore, the asymptotic symmetries are spanned by the same
Lie-algebra valued parameter $\lambda=\lambda(T,Y,\mathcal{E})$ as in section
\ref{Super-BMS} but, since the invariant form has been modified
according to (\ref{CS3}), the global charges acquire a correction, so
that they now read%
\begin{equation}
  \mathcal{Q}[T,Y,\mathcal{E}]=-\frac{k}{4\pi}\int\left[  T \mathcal{M}+Y\left( \mathcal{ J}+\mu
      \mathcal{M}\right)  -2 \mathcal{E}\Psi\right]  d\phi\ . 
\end{equation}
Consequently, once expanded in modes, the Poisson bracket algebra of
the canonical generators are given by the minimal supersymmetric
extension of the BMS$_{3}$ algebra \eqref{Super-BMS-simplest}, but
with an additional central charge,
\begin{equation}
  \label{eq:1}
  c_{1}=\mu\frac{3}{G},\quad\quad c_{2}=\frac{3}{G}. 
\end{equation}

\acknowledgments We thank M.~Ba\~nados, C.~Bunster, O.~Fuentealba, H.~Gonz\'{a}lez,
M.~Henneaux, P.-H.~Lambert, C.~Mart\'{\i}nez, A.~P\'{e}rez, D.~Tempo, and
C.~Troessaert for enlightening discussions. J.M.~and R.T.~wish to
thank the Physique th\'{e}orique et math\'{e}matique group of the
Universit\'{e} Libre de Bruxelles, and the International Solvay
Institutes for the warm hospitality. J.M.~also thanks Conicyt for
financial support. This work is partially funded by the Fondecyt
grants N${^{\circ}}$ 1130658, 1121031. The Centro de Estudios
Cient\'{\i}ficos (CECs) is funded by the Chilean Government through
the Centers of Excellence Base Financing Program of Conicyt. G.B.~is
research director of the Fund for Scientific Research-FNRS Belgium,
while L.D.~is a research fellow of the ``Fonds pour la Formation \`a
la Recherche dans l'Industrie et dans l'Agriculture''-FRIA
Belgium. Their work is supported in part by IISN-Belgium and by
``Communaut\'e fran\c caise de Belgique - Actions de Recherche
Concertées''.

\appendix

\section{Conventions}

\label{conventions}

Our conventions are such that the Levi-Civita symbol fulfills $\epsilon
_{012}=1$, and the tangent space metric $\eta_{ab}$, with $a=0,1,2$, is
off-diagonal, given by%
\[
\eta_{ab}=\left(
\begin{array}
[c]{ccc}%
0 & 1 & 0\\
1 & 0 & 0\\
0 & 0 & 1
\end{array}
\right)  \ .
\]
The three-dimensional $\Gamma$-matrices satisfy the Clifford algebra
$\{\Gamma_{a},\Gamma_{b}\}=2\,\eta_{ab}$, and are chosen as%
\begin{equation}
\Gamma_{0}=\frac{1}{\sqrt{2}}\left(  \sigma_{1}+i\sigma_{2}\right)
,\quad\Gamma_{1}=\frac{1}{\sqrt{2}}\left(  \sigma_{1}-i\sigma_{2}\right)
,\quad\Gamma_{2}=\sigma_{3}\ ,
\end{equation}
where the $\sigma$'s stand for the Pauli matrices%
\[
\sigma_{1}=\left(
\begin{array}
[c]{cc}%
0 & 1\\
1 & 0
\end{array}
\right)  ,\quad\sigma_{2}=\left(
\begin{array}
[c]{cc}%
0 & -i\\
i & 0
\end{array}
\right)  ,\quad\sigma_{3}=\left(
\begin{array}
[c]{cc}%
1 & 0\\
0 & -1
\end{array}
\right)  \ .
\]
As a consequence, they satisfy 
\begin{equation}
  \label{completeness}
  \Gamma_a\Gamma_b=\epsilon_{abc}\Gamma^c+\eta_{ab}\mathbf{1},\quad
  {{\Gamma^a}^\alpha}_\beta{\Gamma_a^\gamma}_\delta=2\delta^\alpha_\delta
\delta^\gamma_\beta-\delta^\alpha_\beta\delta^\gamma_\delta
  \ , 
\end{equation}
where tangent space indices are lowered and raised with $\eta_{ab}$ and
its inverse. 

For a spinor $\psi^{\alpha}$, with $\alpha=+1$, $-1$, we define the Majorana
conjugate as $\bar{\psi}_{\alpha}=C_{\alpha\beta}\psi^{\beta}$, where the charge
conjugation matrix is given by $C=i\sigma_{2}$, that is%
\begin{equation}
C_{\alpha\beta}=\varepsilon_{\alpha\beta}=C^{\alpha\beta}=\left(
\begin{array}
[c]{cc}%
0 & 1\\
-1 & 0
\end{array}
\right)  \ ,
\end{equation}
which satisfies $C^{T}=-C$ and
$C\Gamma_{a}C^{-1}=-(\Gamma_{a})^{T}$. In particular, this implies 
$\overline{\Lambda^{-1}\psi}=\bar\psi\Lambda$ if $\Lambda\in
\mathrm{SL}(2,\mathbf{R})$.


\begin{thebibliography}{100}

\bibitem{Deser:1983tn} 
  S.~Deser, R.~Jackiw and G.~'t Hooft,
  Annals Phys.\  {\bf 152}, 220 (1984).
  
\bibitem{Deser:1983dr} 
  S.~Deser and R.~Jackiw,
  Annals Phys.\  {\bf 153}, 405 (1984).
  

\bibitem{Henneaux:1984ei} 
  M.~Henneaux,
  Phys.\ Rev.\ D {\bf 29}, 2766 (1984).
 
\bibitem{Deser:1985qs} 
  S.~Deser,
  Class.\ Quant.\ Grav.\  {\bf 2}, 489 (1985).
  
  

\bibitem{Witten:1994cga} 
  E.~Witten,
  Int.\ J.\ Mod.\ Phys.\ A {\bf 10}, 1247 (1995)
  [hep-th/9409111].
  
\bibitem{Brown:1986nw} 
  J.~D.~Brown and M.~Henneaux,
  Commun.\ Math.\ Phys.\  {\bf 104}, 207 (1986).
  
\bibitem{Banados:1992wn} 
  M.~Banados, C.~Teitelboim and J.~Zanelli,
  Phys.\ Rev.\ Lett.\  {\bf 69}, 1849 (1992)
  [hep-th/9204099].
  
  
\bibitem{Banados:1993gq} 
  M.~Banados, M.~Henneaux, C.~Teitelboim and J.~Zanelli,
  Phys.\ Rev.\ D {\bf 48}, 1506 (1993)
  [gr-qc/9302012].
  
\bibitem{Navarro-Salas1999} 
  J.~Navarro-Salas and P.~Navarro,
  JHEP {\bf 9905}, 009 (1999)
  [hep-th/9903248].
  
\bibitem{Nakatsu1999} 
  T.~Nakatsu, H.~Umetsu and N.~Yokoi,
  Prog.\ Theor.\ Phys.\  {\bf 102}, 867 (1999)
  [hep-th/9903259].
  
  
\bibitem{Maloney:2007ud} 
  A.~Maloney and E.~Witten,
  JHEP {\bf 1002}, 029 (2010)
  [arXiv:0712.0155 [hep-th]].
  
  
\bibitem{Garbarz:2014kaa} 
  A.~Garbarz and M.~Leston,
  JHEP {\bf 1405}, 141 (2014)
  [arXiv:1403.3367 [hep-th]].
  
\bibitem{Barnich:2014zoa} 
  G.~Barnich and B.~Oblak,
  ``Holographic positive energy theorems in three-dimensional gravity,''
  arXiv:1403.3835 [hep-th].
  
\bibitem{Coussaert:1994jp} 
  O.~Coussaert and M.~Henneaux,
  Phys.\ Rev.\ Lett.\  {\bf 72}, 183 (1994)
  [hep-th/9310194].

 
  
\bibitem{Banados:1998pi} 
  M.~Banados, K.~Bautier, O.~Coussaert, M.~Henneaux and M.~Ortiz,
  Phys.\ Rev.\ D {\bf 58}, 085020 (1998)
  [hep-th/9805165].
  
  
  
\bibitem{Henneaux:1999ib} 
  M.~Henneaux, L.~Maoz and A.~Schwimmer,
  Annals Phys.\  {\bf 282}, 31 (2000)
  [hep-th/9910013].
  
  
\bibitem{Ashtekar1997} 
  A.~Ashtekar, J.~Bicak and B.~G.~Schmidt,
  Phys.\ Rev.\ D {\bf 55}, 669 (1997)
  [gr-qc/9608042].
  
  
\bibitem{Barnich:2006av} 
  G.~Barnich and G.~Compere,
  Class.\ Quant.\ Grav.\  {\bf 24}, F15 (2007), corrigendum: ibid
  24 (2007) 3139
  [gr-qc/0610130].
  
  
  
\bibitem{Barnich:2010eb} 
  G.~Barnich and C.~Troessaert,
  JHEP {\bf 1005}, 062 (2010)
  [arXiv:1001.1541 [hep-th]].


\bibitem{Bagchi:2012cy} 
  A.~Bagchi and R.~Fareghbal,
  JHEP {\bf 1210}, 092 (2012)
  [arXiv:1203.5795 [hep-th]].

  
\bibitem{Barnich:2012aw} 
  G.~Barnich, A.~Gomberoff and H.~A.~Gonzalez,
  Phys.\ Rev.\ D {\bf 86}, 024020 (2012)
  [arXiv:1204.3288 [gr-qc]].
  
  
\bibitem{Gonzalez:2013oaa} 
  H.~A.~Gonzalez, J.~Matulich, M.~Pino and R.~Troncoso,
  JHEP {\bf 1309}, 016 (2013)
  [arXiv:1307.5651 [hep-th]].
  
\bibitem{Costa:2013vza} 
  R.~N.~Caldeira Costa,
  ``Aspects of the zero $\Lambda$ limit in the AdS/CFT correspondence,''
  arXiv:1311.7339 [hep-th].
  
\bibitem{Fareghbal:2013ifa} 
  R.~Fareghbal and A.~Naseh,
  JHEP {\bf 1403}, 005 (2014)
  [arXiv:1312.2109 [hep-th]].
  
\bibitem{Krishnan:2013wta} 
  C.~Krishnan, A.~Raju and S.~Roy,
  JHEP {\bf 1403}, 036 (2014)
  [arXiv:1312.2941 [hep-th]].
  
\bibitem{Cornalba:2002fi} 
  L.~Cornalba and M.~S.~Costa,
  Phys.\ Rev.\ D {\bf 66}, 066001 (2002)
  [hep-th/0203031].
  
\bibitem{Cornalba:2003kd} 
  L.~Cornalba and M.~S.~Costa,
  Fortsch.\ Phys.\  {\bf 52}, 145 (2004)
  [hep-th/0310099].
  
\bibitem{Barnich:2012xq} 
  G.~Barnich,
  JHEP {\bf 1210}, 095 (2012)
  [arXiv:1208.4371 [hep-th]].
  
\bibitem{Bagchi:2012xr} 
  A.~Bagchi, S.~Detournay, R.~Fareghbal and J.~Simon,
  Phys.\ Rev.\ Lett.\  {\bf 110}, 141302 (2013)
  [arXiv:1208.4372 [hep-th]].
  
\bibitem{Strominger:1997eq} 
  A.~Strominger,
  JHEP {\bf 9802}, 009 (1998)
  [hep-th/9712251].
  
\bibitem{Coussaert:1995zp} 
  O.~Coussaert, M.~Henneaux and P.~van Driel,
  Class.\ Quant.\ Grav.\  {\bf 12}, 2961 (1995)
  [gr-qc/9506019].
  
\bibitem{Barnich:2012rz} 
  G.~Barnich, A.~Gomberoff and H.~A.~Gonzalez,
  Phys.\ Rev.\ D {\bf 87}, 124032 (2013)
  [arXiv:1210.0731 [hep-th]].
  
  
\bibitem{Salomonson:1989fw} 
  P.~Salomonson, B.~S.~Skagerstam and A.~Stern,
  Nucl.\ Phys.\ B {\bf 347}, 769 (1990).
  
\bibitem{Barnich:2013yka} 
  G.~Barnich and H.~A.~Gonzalez,
  JHEP {\bf 1305}, 016 (2013)
  [arXiv:1303.1075 [hep-th]].
  

\bibitem{Barnich2014} 
    G.~Barnich and B.~Oblak (2014), to appear

  
\bibitem{Bagchi:2009ke} 
  A.~Bagchi and I.~Mandal,
  Phys.\ Rev.\ D {\bf 80}, 086011 (2009)
  [arXiv:0905.0580 [hep-th]].
  

\bibitem{Mandal:2010gx} 
  I.~Mandal,
  JHEP {\bf 1011}, 018 (2010)
  [arXiv:1003.0209 [hep-th]].
  
  
\bibitem{Sakaguchi:2009de} 
  M.~Sakaguchi,
  J.\ Math.\ Phys.\  {\bf 51}, 042301 (2010)
  [arXiv:0905.0188 [hep-th]].
  
\bibitem{deAzcarraga:2009ch} 
 J. de Azcarraga and J. Lukierski, 
 Phys. \ Lett. \ B678, 411 (2009),
 [arXiv:0905.0141 [math-ph]].

\bibitem{Deser:1982sw} 
  S.~Deser and J.~H.~Kay,
  Phys.\ Lett.\ B {\bf 120}, 97 (1983).

\bibitem{Deser:1984aa} 
  S.~Deser, Quantum Theory of Gravity: Essays in honor of the 60th
  birthday of Bryce S. DeWitt. (Adam Hilger Ltd., 1984), chap. Cosmological Topological Supergravity, p. 374.

  
\bibitem{Marcus:1983hb} 
  N.~Marcus and J.~H.~Schwarz,
  Nucl.\ Phys.\ B {\bf 228}, 145 (1983).
  
\bibitem{Achucarro:1987vz} 
  A.~Achucarro and P.~K.~Townsend,
  Phys.\ Lett.\ B {\bf 180}, 89 (1986).
  
\bibitem{Achucarro:1989gm} 
  A.~Achucarro and P.~K.~Townsend,
  Phys.\ Lett.\ B {\bf 229}, 383 (1989).
  
  
\bibitem{Nishino:1991sr} 
  H.~Nishino and S.~J.~Gates, Jr.,
  Int.\ J.\ Mod.\ Phys.\ A {\bf 8}, 3371 (1993).
  
\bibitem{Howe:1995zm} 
  P.~S.~Howe, J.~M.~Izquierdo, G.~Papadopoulos and P.~K.~Townsend,
  Nucl.\ Phys.\ B {\bf 467}, 183 (1996)
  [hep-th/9505032].
  
  
\bibitem{Banados:1996hi} 
  M.~Banados, R.~Troncoso and J.~Zanelli,
  Phys.\ Rev.\ D {\bf 54}, 2605 (1996)
  [gr-qc/9601003].

  
\bibitem{Giacomini:2006dr} 
  A.~Giacomini, R.~Troncoso and S.~Willison,
  Class.\ Quant.\ Grav.\  {\bf 24}, 2845 (2007)
  [hep-th/0610077].
  
  
  
\bibitem{Gupta:2007th} 
  R.~K.~Gupta and A.~Sen,
  JHEP {\bf 0803}, 015 (2008)
  [arXiv:0710.4177 [hep-th]].
  
  
\bibitem{Fierro:2014lka} 
  O.~Fierro, F.~Izaurieta, P.~Salgado and O.~Valdivia,
  ``(2+1)-dimensional supergravity invariant under the AdS-Lorentz superalgebra,''
  arXiv:1401.3697 [hep-th].
  
\bibitem{Regge:1974zd} 
  T.~Regge and C.~Teitelboim,
  Annals Phys.\  {\bf 88}, 286 (1974).
  
\bibitem{Balachandran:1991dw} 
  A.~P.~Balachandran, G.~Bimonte, K.~S.~Gupta and A.~Stern,
  Int.\ J.\ Mod.\ Phys.\ A {\bf 7}, 4655 (1992)
  [hep-th/9110072].
  
  
  
\bibitem{Banados:1994tn} 
  M.~Banados,
  Phys.\ Rev.\ D {\bf 52}, 5816 (1996)
  [hep-th/9405171].
  
  
\bibitem{Carlip:2005zn} 
  S.~Carlip,
  Class.\ Quant.\ Grav.\  {\bf 22}, R85 (2005)
  [gr-qc/0503022].
  
\bibitem{Perez:2014pya} 
  A.~Perez, D.~Tempo and R.~Troncoso,
  ``Brief review on higher spin black holes'',
  arXiv:1402.1465 [hep-th].
  
  
  
\bibitem{Deser:1977hu} 
  S.~Deser and C.~Teitelboim,
  Phys.\ Rev.\ Lett.\  {\bf 39}, 249 (1977).
  
\bibitem{Abbott:1981ff} 
  L.~F.~Abbott and S.~Deser,
  Nucl.\ Phys.\ B {\bf 195}, 76 (1982).
  
  
\bibitem{Horowitz:1990ap} 
  G.~T.~Horowitz and A.~R.~Steif,
  Phys.\ Lett.\ B {\bf 258}, 91 (1991).
  
\bibitem{Witten:1981mf} 
  E.~Witten,
  Commun.\ Math.\ Phys.\  {\bf 80}, 381 (1981).
  
    
\bibitem{Liu:2002ft} 
  H.~Liu, G.~W.~Moore and N.~Seiberg,
  JHEP {\bf 0206}, 045 (2002)
  [hep-th/0204168].

\end{thebibliography}
\end{document}